\documentclass{article}       
\usepackage{graphicx}
\usepackage[justification=centering]{caption}
\usepackage[center]{subfigure}
\usepackage{float}

\usepackage{amsmath,amssymb,amsthm}
\newtheorem{proposition}{Proposition}
\begin{document}

\title{Stability bounds of a delay visco-elastic rheological model with substrate friction
}
%



\author{Malik A. Dawi $^1$ 
        Jose J. Mu\~noz$^{1,2*}$\\ 
$^1$\small{Laboratori de C\`alcul Num\`eric (LaC\`aN)}\\
\small{ Universitat Polit\`ecnica de Catalunya, Barcelona, Spain}\\
$^2$\small{Dept. of Mathematics, Universitat Polit\`ecnica de Catalunya, Barcelona, Spain}\\
\small{Centre Internacional de M\`etodes Num\`erics en Enginyeria         (CIMNE), Barcelona, Spain.}\\
\small{              {\tt j.munoz@upc.edu}           }
}


\maketitle


\begin{abstract}
Cells and tissues exhibit oscillatory deformations during remodelling, migration or embryogenesis. Although it has been shown that these  oscillations correlate with cell biochemical signalling, it is yet unclear the role of these oscillations in triggering drastic cell reorganisation events or instabilities, and the coupling of this oscillatory response with tested visco-elastic properties. 

We here present a rheological model that incorporates elastic, viscous and frictional components, and that is able to generate oscillatory response through a delay adaptive process of the rest-length. We analyse its stability properties as a function of the model parameters and deduce analytical bounds of the stable domain. While increasing values of the delay and remodelling rate render the model unstable, we also show that increasing friction with the substrate destabilise the oscillatory response. Furthermore, we numerically verify that the extension of the model to non-linear strain measures is able to generate sustained oscillations that alternate between stable and unstable regions.

\textbf{keywords:}Oscillations, Delay differential equations, Visco-elasticity,  friction , stability, rheology, cells.
\end{abstract}


\section{Introduction}\label{s:intro}

Oscillatory cell deformations are ubiquitous and have been quantified \emph{in vitro} \cite{petrolli19,peyret18} and \emph{in vivo}, for instance in the segmented clock of mice \cite{yoshiokakobayashi20} or during \emph{Drosophila} fly dorsal closure \cite{solon09}. These oscillations have been associated to biochemical dynamics \cite{kaouri19}, signalling delays \cite{petrungaro19} or myosin concentration fluctuations \cite{dierkes14}. We here present a rheological model that explicitly incorporates the delay between the cell length adaptation  and the current stretch.

Time delay has been included in numerous models in biology, with applications in biochemical negative feedback \cite{lapytsko16}, cell growth and division  \cite{alarcon14,gyllenberg87}, or cell maturation \cite{getto19}, but are less common in biomechanics. In our case we introduce this delay in an evolution law of the cell or tissue rest-length. Such models with varying rest-length have been applied to stress relaxation \cite{khalilgharibi19}, morphogenesis \cite{clement17}, cortical mechanics \cite{doubrovinski17}, or endocytosis \cite{cavanaugh20}. They have the advantage of including a measurable quantity, the rest-length \cite{wyatt20}, and also furnishing the observed visco-elastic response. We will here adapt these models and include the delay response in conjunction with frictional or adhesive forces from the environment or substrate.

Our visco-elastic model mimics the standard linear solid, but expressed in terms of delay rest-length changes, which provides the oscillatory character of the deformation. The stability of such system has been described in \cite{munoz18} or in \cite{borja20} for planar frictionless dynamics of monolayers.  We here extend such analysis to a frictional substrate, and deduce the stability conditions as a function of viscous, stiffness and friction parameters. 

The stability analysis is usually carried out through the inspection of the characteristic equation \cite{asl03,smith11}, or semi-discretisation methods \cite{insperger02,sykora19}. We resort to the former method, and by analysing the associated Lambert function \cite{shinozaki06,corless96}, we deduce strict and simple bounds of the stability region. We compare our analysis with some numerical solutions of the Delay Differential Equations (DDEs). 

The article is organised as follows. We describe the visco-elastic model in Section \ref{s:model} together with the delay evolution law of the rest-length. In Section \ref{s:lstab} the stability of a linear model is analysed, and some bounds as a function of the model parameters are given. A non-linear extension is presented in Section \ref{s:nlstab}, which is solved numerically and is analysed with the help  of the results obtained in the linearised model. Our findings are finally discussed in the Conclusions section.

\section{Visco-elastic model with delay} \label{s:model}

\begin{figure*}
\centering
  \includegraphics[width=0.4\textwidth]{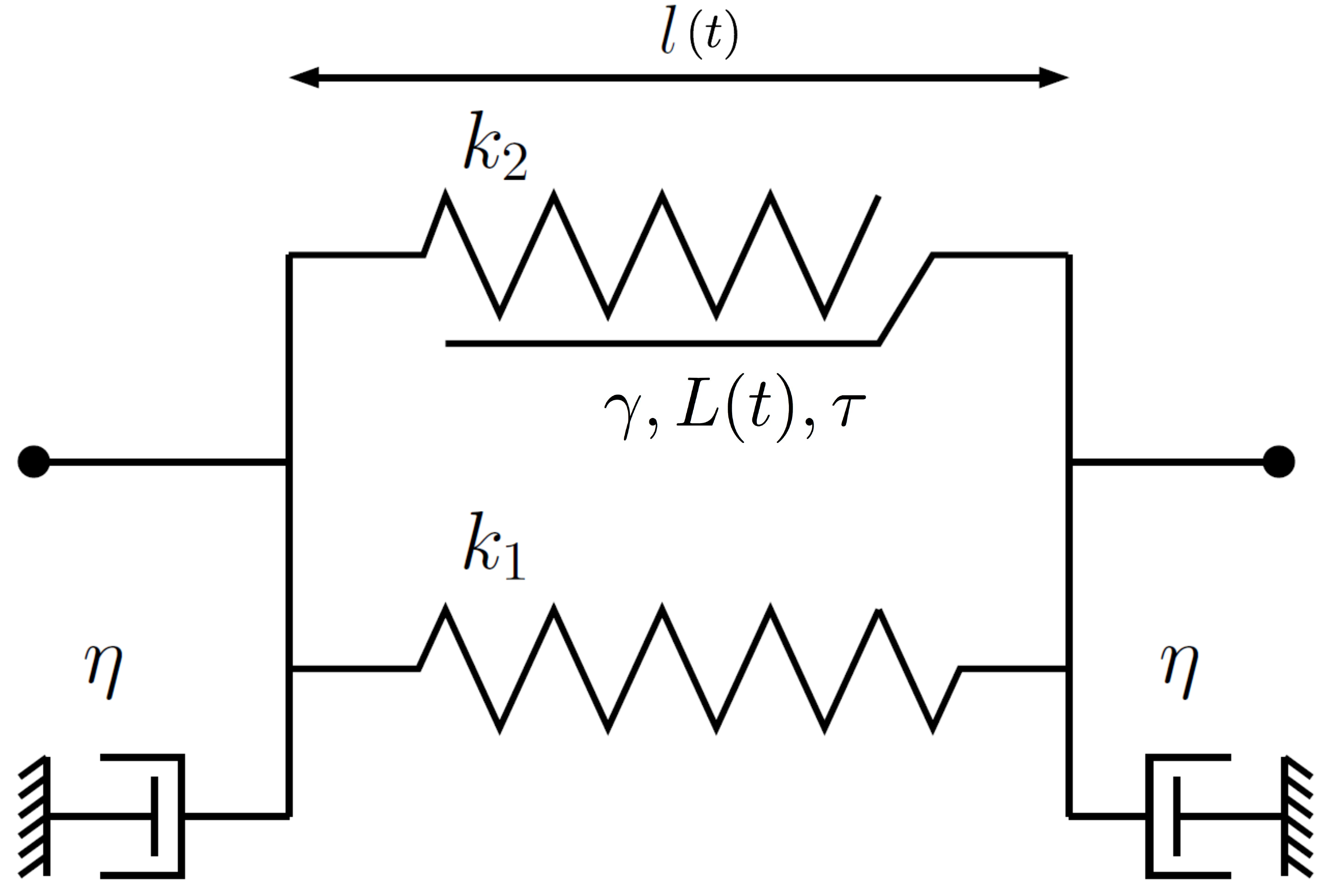}
\caption{Schematic view of 1-dimensional model, illustrating both elastic and visco-elastic branches with dissipative friction.}
\label{f:rheology}       
\end{figure*}

We consider a material rheology that mimics the solid standard mode: a purely elastic stress $\sigma^e$ in parallel with a visco-elastic stress $\sigma^v$. Figure \ref{f:rheology} shows schematically the two branches. We assume a one-dimensional domain $D=\left[0, l(t)\right]$, with $l(t)$ a time dependent apparent (measurable) length of the domain. 

The total stress $\sigma$ in $D$ is given by the sum of elastic and viscoelastic contributions,
\begin{align*}
\sigma=\sigma^e + \sigma^v,
\end{align*}
where each stress component is given by $\sigma^e=k_1 \varepsilon(l(t), l_0)$ and $\sigma^v=k_2\varepsilon^e(l(t), L(t))$, with $k_1$ and $k_2$ the associated stiffness parameters. The strain measures $\varepsilon(l(t), l_0)$ and $\varepsilon^e(l(t), L(t))$ will be detailed in the next sections for the linear and non-linear models. As yet we mention that they depend, in addition to $l(t)$, on the initial length $l_0=l(0)$ and the rest-length $L(t)$ of the visco-elastic branch. This rest-length can be  interpreted as an internal variable, whose evolution mimics the viscous response of Maxwell models \cite{munoz13b}.

More specifically, $L(t)$ changes according to the following evolution law
\begin{equation}\label{e:evolL}
    \dot L(t)=\gamma ( l(t-\tau)-L(t-\tau)), t> 0.
\end{equation} 

Henceforth we denote by a superimposed dot the time derivatives, i.e. $\dot{(\bullet)}=d(\bullet)/dt$. Parameter $\gamma>0$ is the \emph{remodelling rate}, which measures the rate at which the cell adapts its length to the difference $l(t-\tau)-L(t-\tau)$. We have introduced the delay parameter $\tau\geq 0$ which aims at mimicking the measured time-lag between the chemical signalling and the internal mechanical remodelling in the cell, as measured in different systems such as Drosophila dorsal closure \cite{dierkes14} or in wound healing \cite{zuluetacoarasa18}, and which in these cases  is in the order of a few minutes.  

We also include in our model a viscous friction $\sigma_\eta$ with the external substrate or environment, and given by an external force $\sigma_\eta(t)=-\eta\dot l(t)$, with $\eta\ge0$ a viscous coefficient (see Figure \ref{f:rheology}). In total, the balance law, $\sigma_\eta = \sigma^e + \sigma^v$ reads in our case
\begin{equation}\label{e:equil}
    -\eta\dot l(t)=k_1\varepsilon^e(l(t), l_0) + k_2\varepsilon ^e (l(t), L(t)),\ t>0, 
\end{equation}
which should be solved together with the evolution law in \eqref{e:evolL}. Due to the presence of the delay $\tau$, initial conditions must be specified for $t\in[-\tau,0]$. For simplicity, we assume constant values
\begin{align}\label{e:ic}
l(t)=l_0,\ t\in [-\tau, 0], \\
L(t)=L_0,\ t\in [-\tau, 0],
\end{align}
with $l_0$ and $L_0$ given constants. In the next sections we will analyse the stability and oscillatory regime of the system of Delay Differential Equations (DDE) for linear and non-linear definitions of the strain measures $\varepsilon$ and $\varepsilon^e$. 


\section{Stability analysis of linear model}\label{s:lstab}

\subsection{Characteristic equations and analytical bounds}

In order to ease the stability analysis, we assume here linear definitions of the strain measures:
\begin{align*}
\begin{aligned}
\varepsilon(l(t), l_0)&=l(t)-l_0,\\
\varepsilon^e(l(t), L(t))&=l(t)-L(t).
\end{aligned}
\end{align*}

Inserting these expression into the balance equation \eqref{e:equil}, the set of DDE turn into the following form:
\begin{align} \label{e:meq}
    -\eta \dot{l}(t) & = k_1 \left(l(t)-l_0\right) + k_2 \left(l(t)-L(t)\right) , &t>0\\
        \dot L(t)&=\gamma ( l(t-\tau)-L(t-\tau)), & t>0
\end{align} 
with the initial conditions in \eqref{e:ic}. The coupled system of DDE can be written in a compact form as 
\begin{equation} \label{e:systemDF}
    \dot{\mathcal{L}}(t)+\mathbf A \mathcal{L}(t)+\mathbf{B}\mathcal{L}(t-\tau)+\mathbf c=\mathbf 0, t>0,
\end{equation}
with  
\begin{equation*}
    \mathcal L(t)=\left\{\begin{array}{c}
    l(t) \\
    L(t)
    \end{array}\right\}\ ; \ 
    \mathbf A=\left[\begin{array}{cc}
      \frac{ k_1+k_2 }{\eta}& -\frac{k_2 }{\eta}\\
    0 & 0\end{array}\right]\ ; \ 
    \mathbf B=\left[\begin{array}{cc}
     0 & 0\\
      -\gamma & \gamma \end{array}\right]\ ; \ 
      \mathbf c=\left\{\begin{array}{c} \frac{k_1 l_0}{\eta} \\ 0 \end{array}   \right\}.
\end{equation*}

Generally, the solution of the coupled system of DDE in \eqref{e:systemDF} is characterized qualitatively (e.g. asymptotic, synchronous, oscillatory) by the exponents or the roots of the characteristic function \cite{erneux09,smith11}. In order to obtain this characteristic function, one might search for a solution in the form,
\begin{equation}\label{e:solution}
    \mathcal{L}(t)=\sum_{i} e^{m_it}\mathcal{L}_i  + \mathcal{L}_0,
\end{equation}
where  $\mathcal L_0$ and $\mathcal{L}_i$ are constant vectors that depend on the chosen initial values, and $m_i \in \mathbb{C} $  are the characteristic exponents. Clearly if all the exponent have negative real parts, i.e.  $Re(m_i)<0$, the solution is asymptotically stable with time. Substituting Eq. \eqref{e:solution} into Eq. \eqref{e:systemDF} gives for each term in the summation
\begin{equation*}
    \left( m_i \mathbf{I} + \mathbf{A}+\mathbf{B} e^{-m_i \tau} \right) \mathcal{L}_i = \mathbf 0.
\end{equation*}

We remark that the above linear transformation must hold regardless of the initial conditions, that is to say, the determinant must always vanish. This allows us to express the characteristic function of the system as the determinant of the above matrix, which gives
\begin{equation} \label{e:characteristic_function}
  f(m):=  m^2+\gamma m e^{-m\tau} + \frac{k_1+k_2}{\eta} m + \frac{\gamma k_1}{\eta} e^{-m\tau} =0.
\end{equation}

We decompose the characteristic function to real and imaginary parts by substituting $m=\alpha+i \beta$ and then separating each part, leading to the following non-linear system of equations, 
\begin{equation}
\begin{aligned} \label{ReIm}
    \text{Re}\; f(m) &= \alpha^2-\beta^2+ \frac{k_1+k_2}{\eta} \alpha+\gamma e^{-\alpha \tau}\left( \left(\alpha+\frac{k_1}{\eta}\right) \cos(\beta \tau)+ \beta \sin(\beta \tau) \right), \\
    \text{Im}\;f(m)& =2\alpha\beta+\frac{k_1+k_2}{\eta} \beta+ \gamma e^{-\alpha \tau} \left( \beta \cos(\beta \tau) - \left(\alpha + \frac{k_1}{\eta}\right) \sin(\beta \tau) \right). 
\end{aligned}
\end{equation}

The stability regions in the parameters space are defined by the borders where the number of unstable exponents changes, which means, at least one characteristic exponents crosses the imaginary axes from left to right. In such case Eq. \eqref{ReIm} will have at least one solution with positive $\alpha$. 

\begin{figure}[!htb]
 \centering 
 \captionsetup{justification=centering}
 \centering 
\subfigure[]{\label{(k1,eta)}\includegraphics[width=0.45\textwidth]{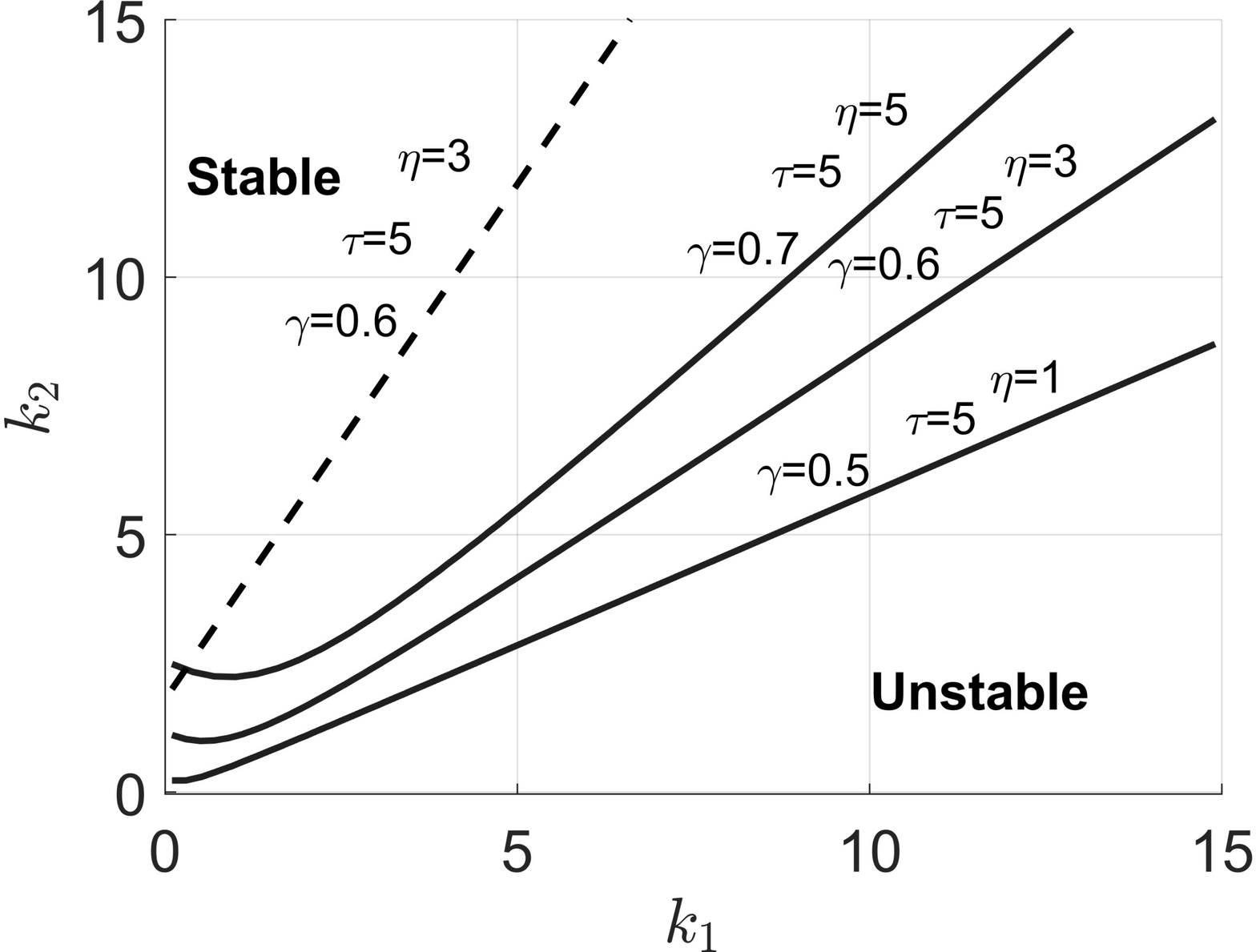}}
\subfigure[]{\label{ph12}\includegraphics[width=0.45\textwidth]{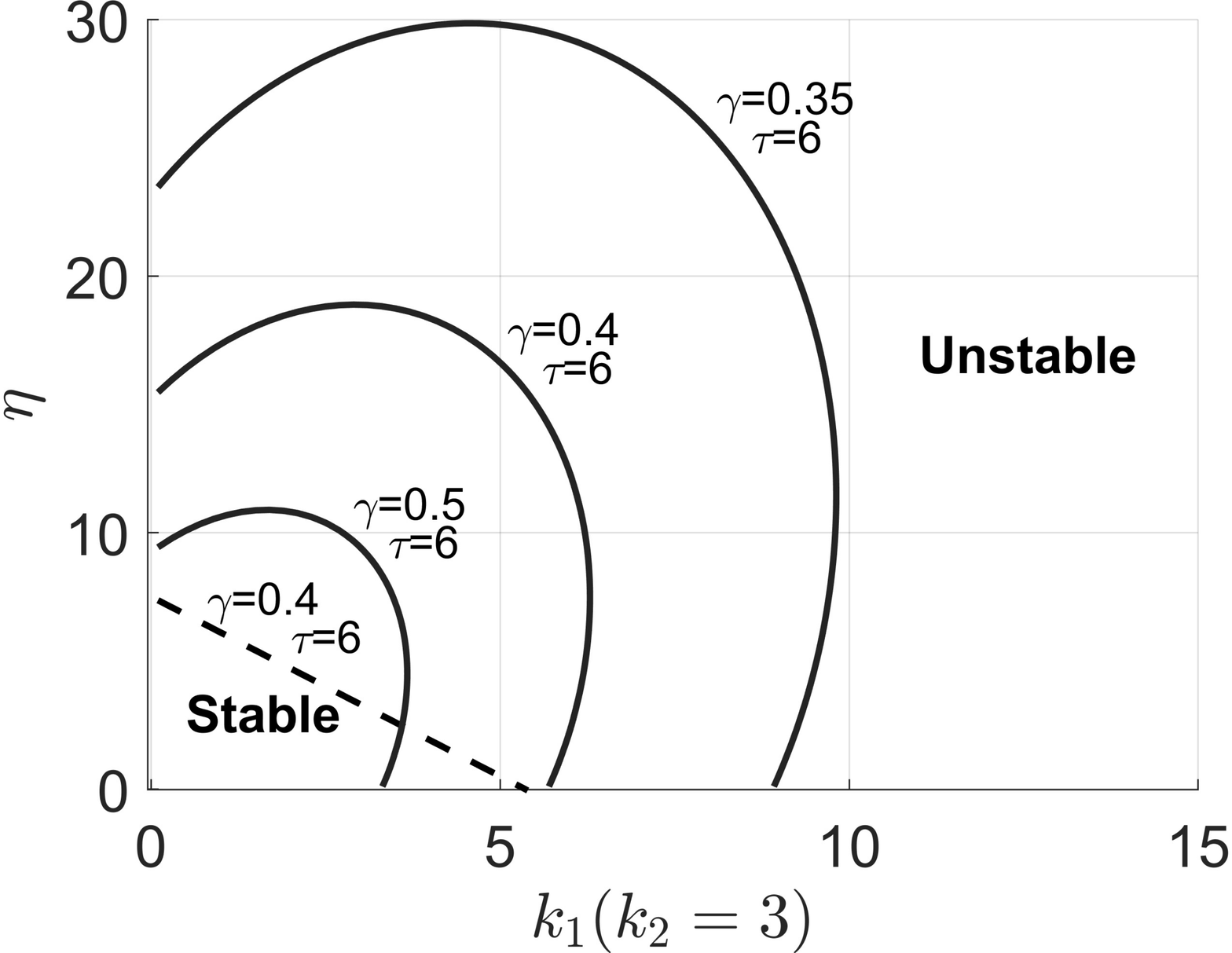}}
\subfigure[]{\label{ph21}\includegraphics[width=0.45\textwidth]{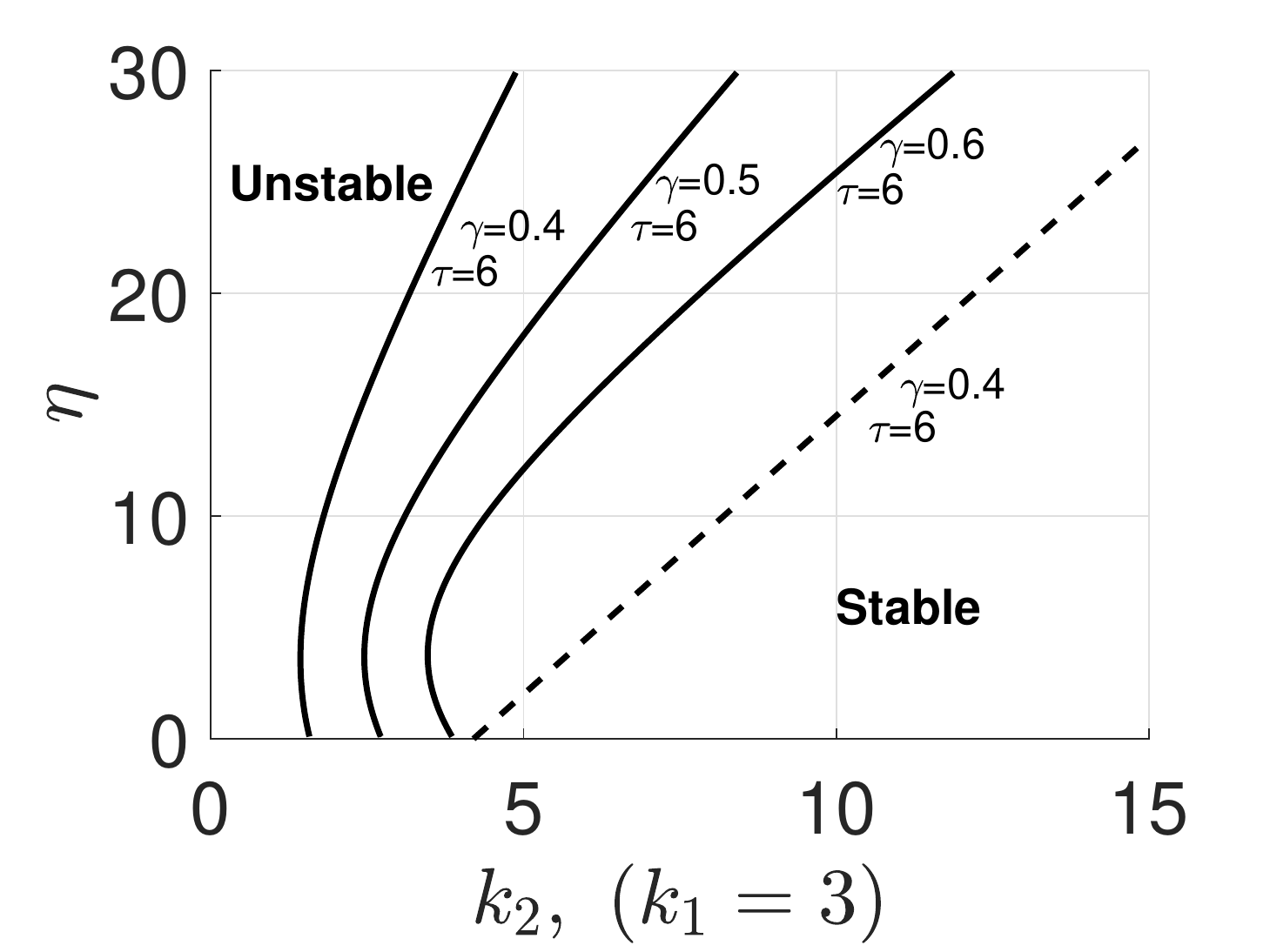}}
\subfigure[]{\label{ph22}\includegraphics[width=0.45\textwidth]{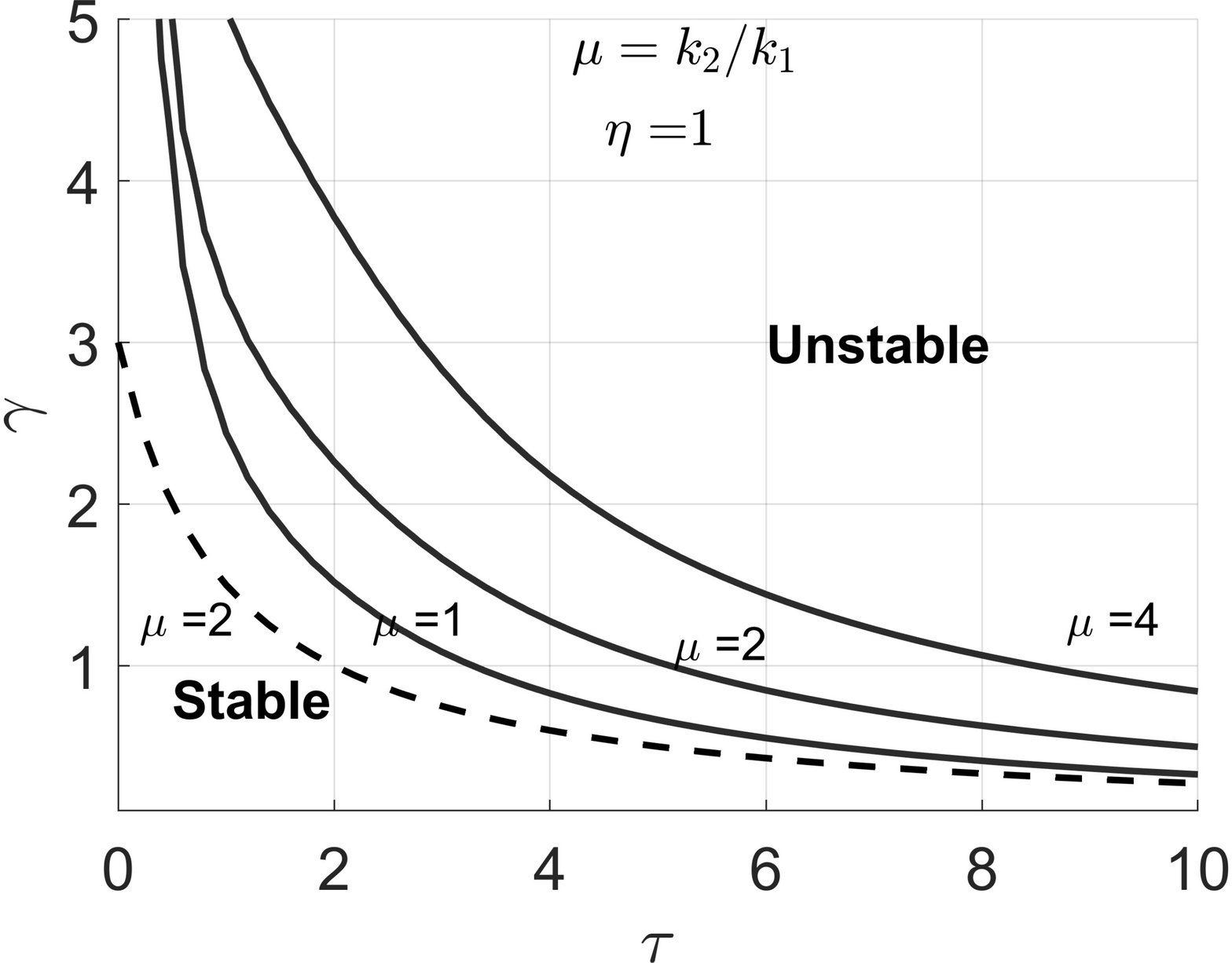}}
\caption{Phase diagrams for different pairs of material parameters. (a) Plane  $(k_1, k_2)$, (b) plane $(k_1,\eta)$, (c) plane $(k_2,\eta)$ and (d) plane $(\tau,\gamma)$. The curves show stability borders for different values of the off-plane parameters. Continuous lines are obtained with the numerical solution of Eq. \eqref{ReIm}. Dashed lines represent the sufficient stability condition in Eq. \eqref{condition}. The regions which are labeled as \textit{stable} are those with negative values for $\alpha$ and those label as  \textit{unstable} indicate the regions with at least a single positive $\alpha$.}
\label{f:StabilitySpace}
\end{figure}

Here, we have constructed the phase diagram by solving the system in Eq. \eqref{ReIm} numerically while monitoring the values of $\alpha$ (see Fig. \ref{f:StabilitySpace}). If there is at least one root with a positive $\alpha$ the solution was considered unstable.

With the aim of furnishing a practical bound for detecting stable solutions, we also give the following result:
\begin{proposition} \label{proposition1}
The solution of the system of delay differential equations in Eq. \eqref{e:systemDF} with initial conditions in Eq. \eqref{e:ic} is stable as long as, 
\begin{equation} \label{condition}
   k_1+k_2- \gamma \eta  -   k_1 \gamma \tau  > 0.
\end{equation}
\end{proposition}

\textbf{Proof}. Condition (\ref{condition}) is derived resorting to the results in \cite{stepan89}, and analysing the so-called \textit{D-curves} defined as, 
\begin{align}
    R(\omega) := \text{Re}\; f(i \omega) &= -\omega^2+\gamma \left( \frac{k_1}{\eta}\cos(\omega \tau)+ \omega\sin(\omega \tau) \right) \\
    S(\omega) := \text{Im}\;f(i \omega)& =\frac{k_1+k_2}{\eta} \omega+ \gamma  \left( \omega \cos(\omega \tau) -   \frac{k_1}{\eta} \sin(\omega \tau) \right) 
\end{align}
with $\omega \in [0,+\infty)$. The functions $R(\omega)$ and $S(\omega)$ provide infinite parametric curves that mark the region with constant number of unstable characteristic exponents. In particular, we resort to \textit{Theorem 2.19} in \cite{stepan89}, which indicates that the zeros of Eq. \eqref{e:characteristic_function} have no real positive parts if and only if,  
\begin{equation} \label{StabilityCondition1}
    S(\rho_k)  \neq 0 \quad k=1,..,r, 
\end{equation}
and
\begin{equation} \label{StabilityCondition2}
    \sum_{k=1}^r (-1)^k S(\rho_k)=-1, 
\end{equation}
where $\rho_1 \geq ...\geq \rho_r \geq 0 $ are the non-negative roots of $R(\omega)$,  with $r$ being an odd number. Moreover, we introduce a polynomial $S^-(\omega)$ which defines a lower bound for the function $S(\omega)$ such that, 
\begin{equation} \label{StabilityCondition3}
    0 < S^-(\omega)\leq S(\omega) \quad \quad  \text{for} \quad  \omega \in (0,+\infty).
\end{equation}

In case that $S(\omega)$ satisfies the stability conditions in Eq. \eqref{StabilityCondition1} and  \eqref{StabilityCondition2}, $S^-(\omega)$ will also satisfy them by construction. An adequate choice for the polynomial $S^-(\omega)$ can be obtained by exploiting the following inequalities, 
\begin{equation*}
    \cos(\omega \tau) \geq -1, \quad -\sin(\omega \tau) \geq - \omega \tau \quad \quad \text{for} \quad \omega \in (0,+\infty)
\end{equation*}
which lead to,
\begin{equation*}
    S^-(\omega)  = \Big( \frac{k_1+k_2}{\eta}- \gamma  -   \frac{k_1}{\eta} \gamma \tau \Big) \omega. 
\end{equation*}

Since $\omega > 0$, the condition in Eq. \eqref{StabilityCondition3} is satisfied as long as
\begin{equation*}
k_1+k_2- \gamma \eta  -   k_1 \gamma \tau  > 0.  \quad \quad \qed
\end{equation*}

We point out that the main benefit of Proposition \ref{proposition1} is that it counts in the whole space of system parameters, giving the opportunity to cross check the stability taking into account the relative variations of system parameters. In the phase diagrams in the parametric space, condition \eqref{condition} is indicated by the dashed lines in Fig. \ref{f:StabilitySpace}. As it can be observed, it indicates stability regions that are smaller then those obtained by solving numerically Eq. \eqref{ReIm}. These plots emphasise the fact that although  the bound in Eq. \eqref{StabilityCondition3} does not provide a necessary condition, it provides a useful sufficient stability condition. 

We remark also two salient conclusion from the expression in the bound, which are also confirmed in the phase diagrams: increasing values of $\gamma\tau$ have an unstable effect in the lengths $l(t)$ and $L(t)$, as previously encountered in other models \cite{munoz18}, while decreasing values of $\eta$ may render the oscillations stable. This is an unexpected result, since increasing viscosity has in general a stabilising or damping effects in mechanics. This can be explained by highlighting the retardation or delay that viscosity entails in the stress response, similar to an increase of $\tau$.

\subsection{Numerical simulations}

In order to verify the obtained stability limits, we have preformed some numerical tests considering the one-dimensional model presented in Fig. \ref{f:rheology}. The test mimics a previous compression state that is given by the following initial conditions,
\begin{align}\label{e:ICResult}
    l(t)=L(t)=1, &\tau < t \leq 0 \\
    l(-\tau)=0.9, L(-\tau)=1. 
\end{align}

In order to compare our results with previous values in the literature and with more general boundary conditions, we will also test different prescribed values of $l(t)$ and additional external forces. Indeed, in the presence of a constant external force $f$, the equilibrium equation in \eqref{e:equil} reads,
\begin{align}\label{e:equilf}
-\eta\dot l(t)+f&=k_1\left(l(t)-l_0\right)+k_2\left(l(t)-L(t)\right)\\
\dot L(t)&=\gamma\left(l(t-\tau)-L(t-\tau)\right)
\end{align}

\subsubsection{Unloaded free conditions} 
 
A backward Euler implicit time discretisation of equations in \eqref{e:equilf} yields the following set of equations, which are computed sequentially, 
\begin{equation} \label{DDENumerical}
\begin{split}
    L_{n+1} & =\Delta t \gamma (l_{n-
    \tau} - L_{n-\tau}) \\
    l_{n+1} & =\frac{1}{(\eta / \Delta t + k_1+k_2)} \left( \frac{\eta}{\Delta t} l_n + f_{n+1}+k_1 L_0 + k_2 L_{n+1} \right)
\end{split}
\end{equation}

\begin{figure}[!htb]
 \centering 
 \captionsetup{justification=centering}
 \centering 
\subfigure[Model parameters: $k_1=2$, $k_2=3$,  $\eta=8$, $\gamma=0.5$, $\tau=6$]{\label{NumericalSolutionStable}\includegraphics[width=50mm]{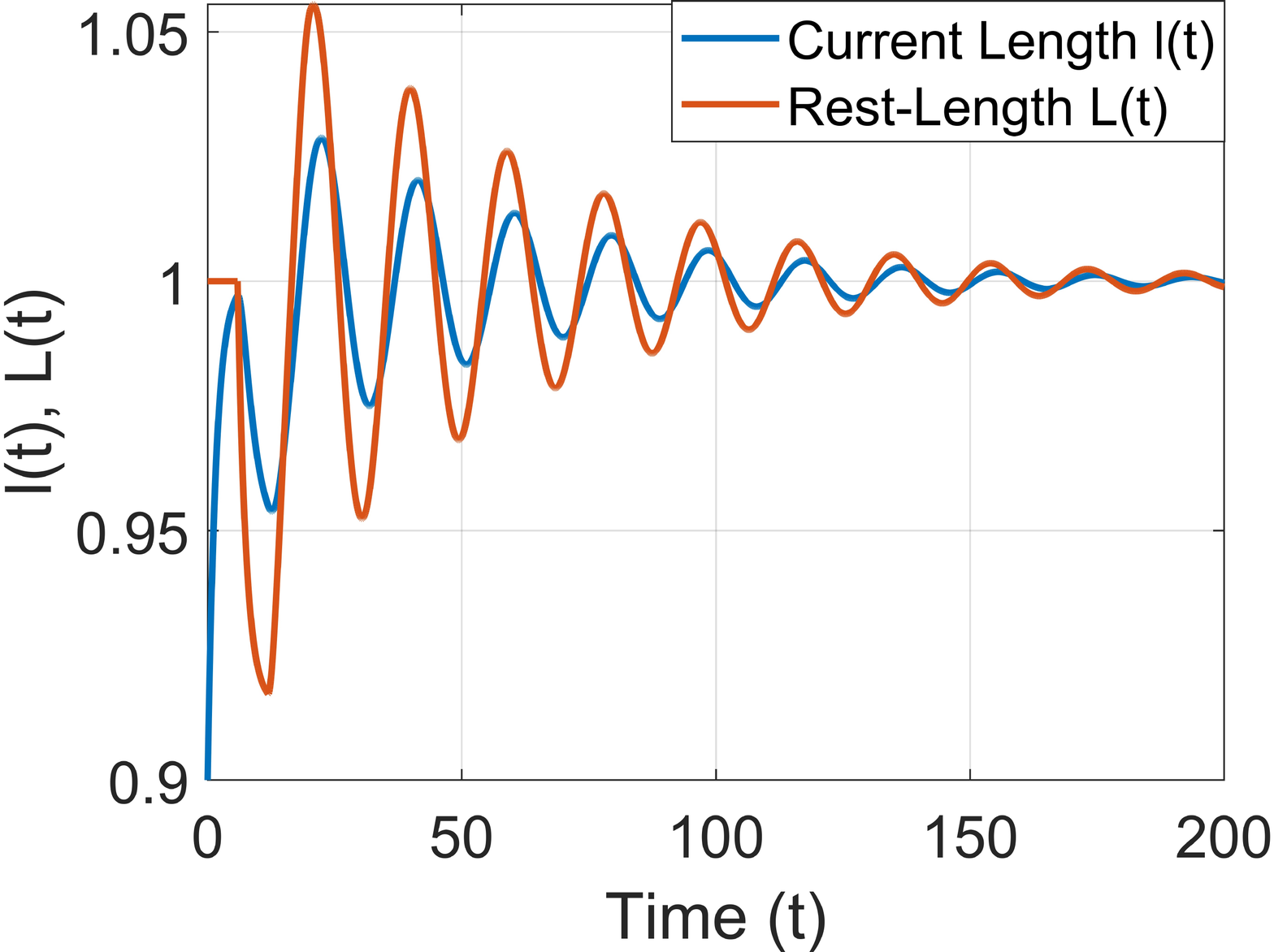}}
\hspace{1cm}
\subfigure[Model parameters: $k_1=3$, $k_2=2$,  $\eta=8$, $\gamma=0.5$, $\tau=6$]{\label{NumericalSolutionUnstable}\includegraphics[width=50mm]{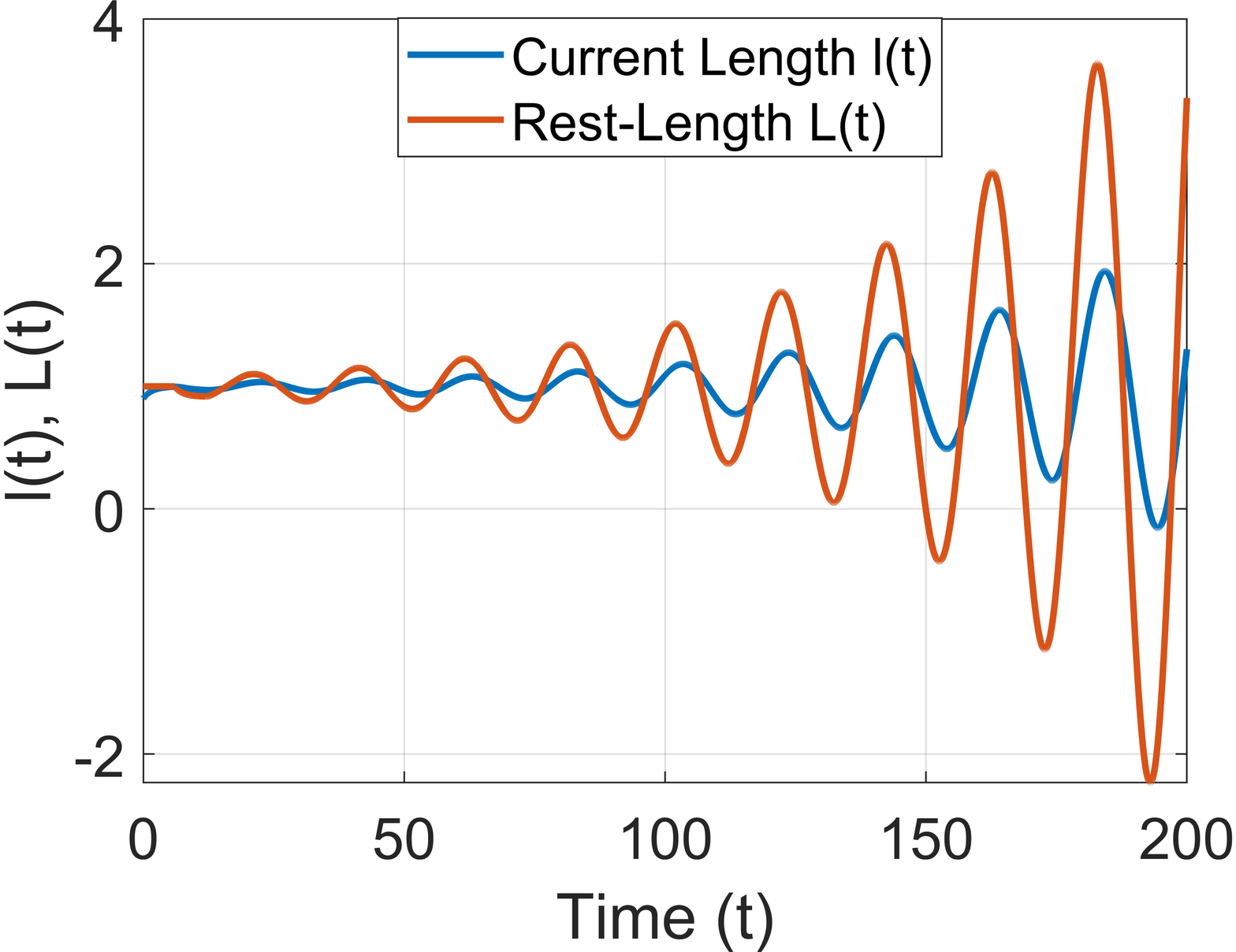}}
\caption{Time evolution of current length and rest-length for free unloaded conditions. (a) Parameters belonging to the stable domain. (b) Choice of parameters that lie outside of the stable domain.   }
\label{f:NumericalSolution}
\end{figure}

We here consider the case $f_n=0, n=0,1,2 \ldots, 200/\Delta t $ and $\Delta t=0.01$, which is found sufficiently accurate when being compared with smaller values. The resulting evolution of $l_n$ and $L_n$ is consistent with the stability analysis of the previous section. The presence of the delay $\tau>0$ produces oscillatory solutions for $l$ and $L$, as it can be seen in Fig. \ref{f:NumericalSolution}. The stability of these oscillations depends on the model parameters as indicated in the stability diagrams in Fig. \ref{f:StabilitySpace}. The first case in Fig. \ref{f:NumericalSolution}a corresponds to stable oscillations, with parameters inside the stability domain, while the second case in Fig, \ref{f:NumericalSolution}b yields unstable oscillations, with parameters that exceed the stability limits.     

\subsubsection{Prescribed deformation}

We here choose a constant value of the apparent length $l(t)$, with an initial discontinuity:
\begin{align*}
    L(t)=L_0=1,& \ -\tau \le t \leq 0, \\
    l(-\tau)=0.9 , l=l_0=1, &\ -\tau < t.
\end{align*}

In this case, $\dot l(t)=0, t>0$, so the the first differential gives us a reaction force term equal to $k_2(l_0-L(t))$, while the DDE reads
\begin{align*}
\dot L&=\gamma (l_0-L(t-\tau)).
\end{align*}

\begin{figure}[!htb]
 \centering 
 \captionsetup{justification=centering}
 \centering 
\subfigure[Model parameters:  $\gamma=0.35$, $\tau=4$]{\label{Stable_l_constant}\includegraphics[width=0.49\linewidth]{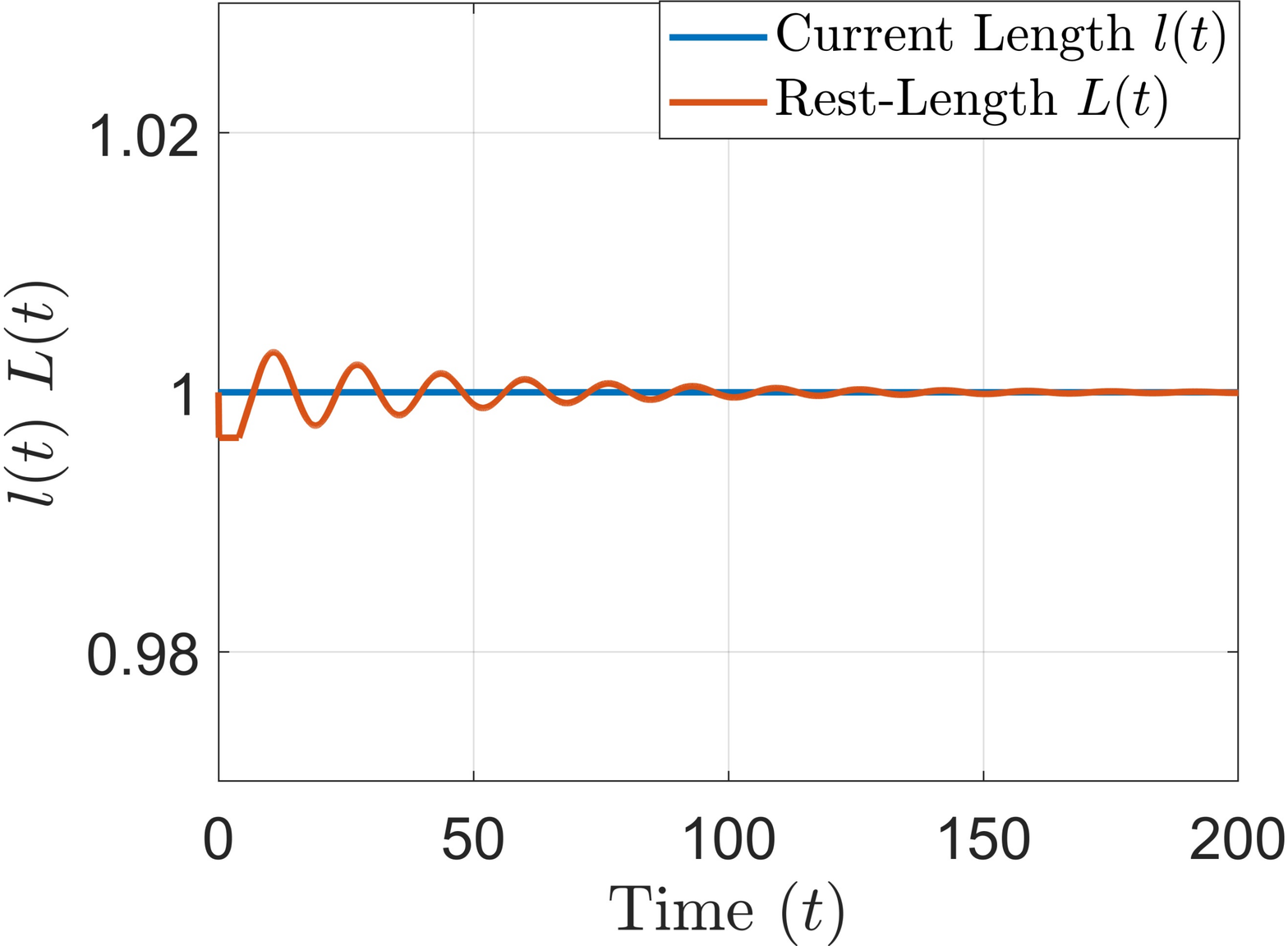}}
\centering
\subfigure[Model parameters: $\gamma=0.35$, $\tau=5$]{\label{Unstable_l_constant}\includegraphics[width=0.49\linewidth]{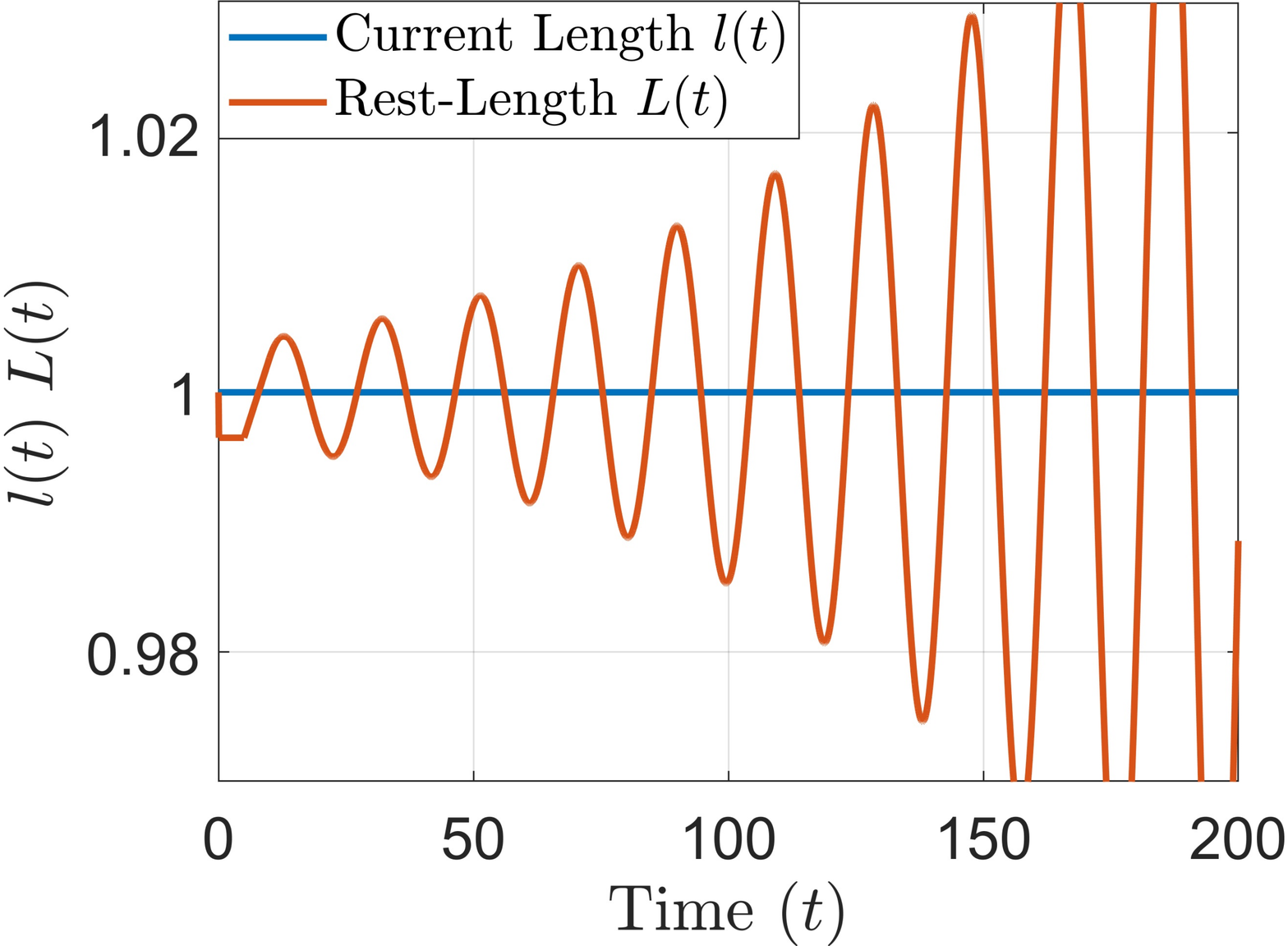}}
\caption{The evolution of the rest-length with fixed values for the apparent length $l(t)$. The stability is in this case identical to the friction-less models \cite{munoz18}: (a) Oscillatory solution when $\tau \gamma > \frac{1}{e}$, (b) unstable solution arise whenever $\tau \gamma > \frac{\pi}{2}$.}
\label{f:prescribed_l}
\end{figure}

This DDE (or equivalent forms) has been extensively studied \cite{smith11,munoz18}, and is known to yield oscillatory values of rest-length $L(t)$ whenever $\gamma\tau>\frac{1}{e}$, and unstable oscillations whenever $\gamma\tau>\frac{\pi}{2}$. This has been confirmed by the numerical simulations in Fig. \ref{f:prescribed_l}.

\subsubsection{Prescribed forces}

\begin{figure}[!htb]
 \centering 
 \captionsetup{justification=centering}
 \centering 
\subfigure[Model parameters: $k_1=1$, $k_2=1$,  $\eta=1$, $\gamma=0.5$, $\tau=6$]{\label{Stable_with_force}\includegraphics[width=56mm]{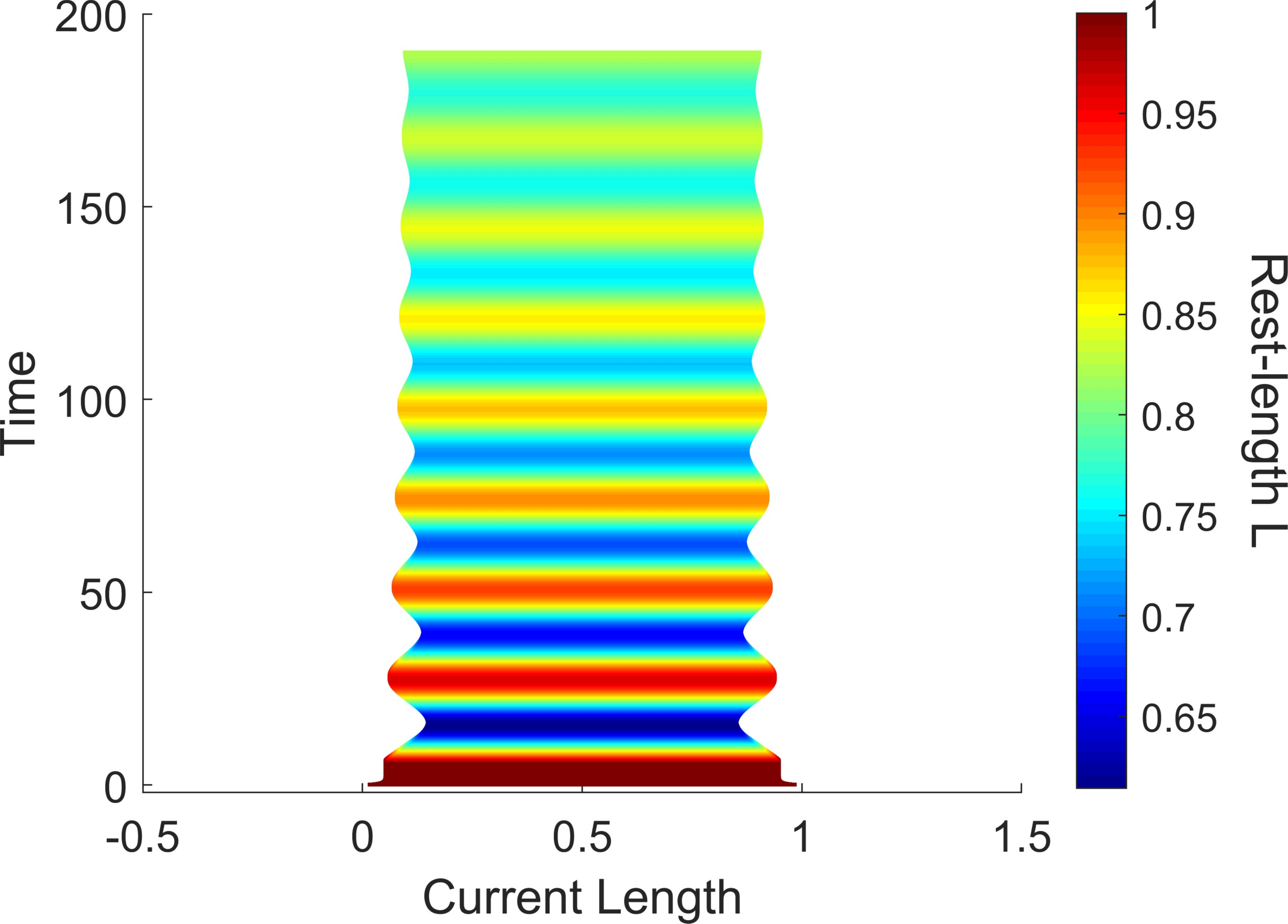}}
\centering
\hspace{.5cm}
\subfigure[Model parameters: $k_1=1$, $k_2=1$,  $\eta=3$, $\gamma=0.6$, $\tau=6$]{\label{Unstable_with_force}\includegraphics[width=56mm]{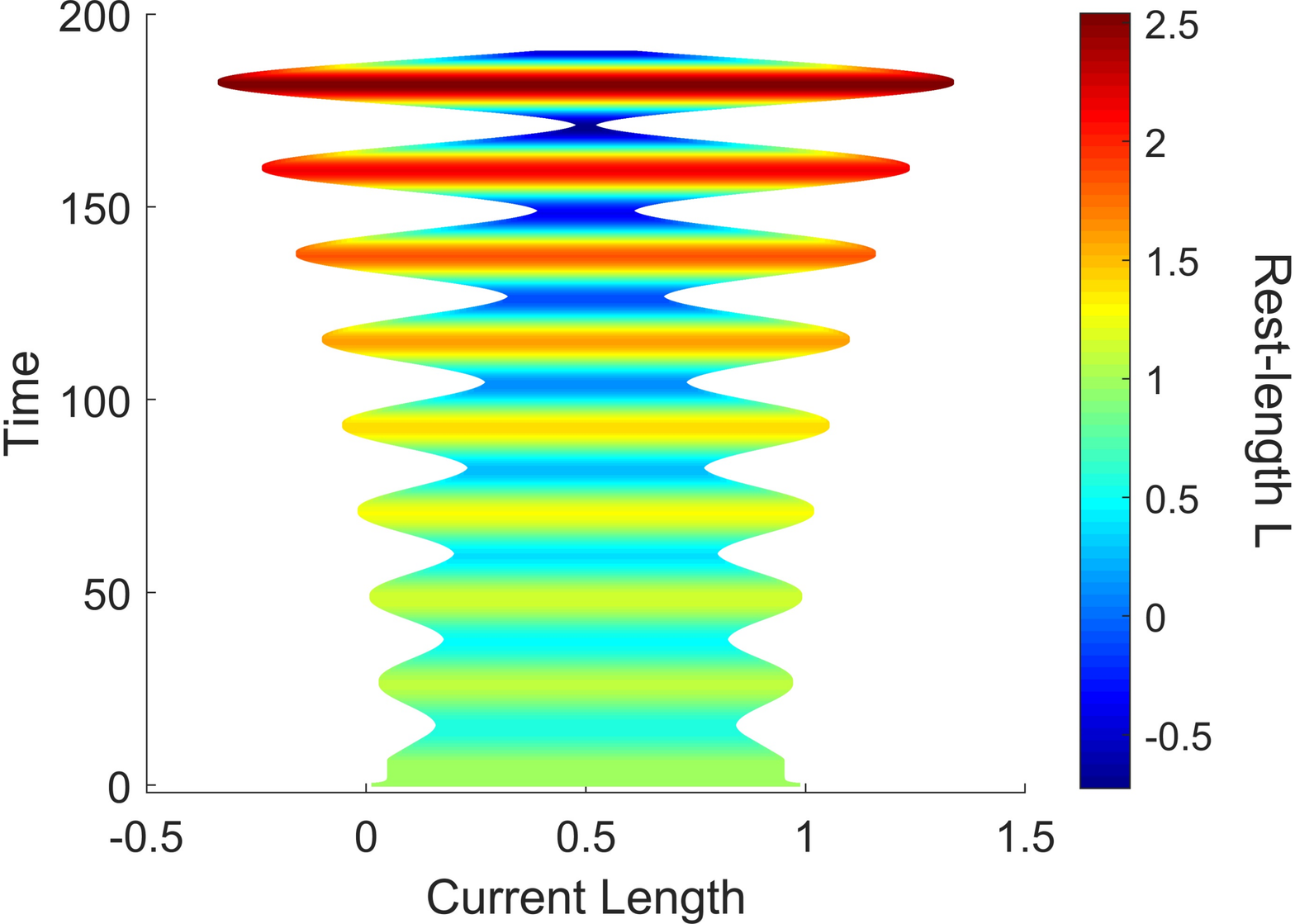}}
\caption{ The evolution of the current length and the rest-length (color map) with prescribed compression forces $f$ ($f(x=0)=0.2, \quad f(x=1)=-0.2$). (a) the solution inside the stability domain. (b) the time evolution as the stability limit is exceeded.}
\label{f:Prescribed_forces}
\end{figure}

We now impose and external force $f=0.2$. Since this value only affects the value of the vector $\mathbf c$ in Eq. \eqref{e:meq}, the stability is consequently unaffected by the value of $f$. The plots in Fig. \ref{f:Prescribed_forces} confirm this fact. These plots show the apparent length as a function of time, while the rest-length is shown as the contourplot on the varying domain $x\in [0,l(t)]$.

\section{Extension to non-linear: strain--based model} \label{s:nlstab}

We now use a non-dimensional definition of the strains
\begin{align*}
\varepsilon(l(t), l_0)&=\frac{l(t)-l_0}{l_0},\\
\varepsilon^e(l(t), L(t))&=\frac{l(t)-L(t)}{L(t)}.
\end{align*}

While this is a more common strain measure, with non-dimensional values, these expressions, when inserted into the equilibrium equations in \eqref{e:equil}  yield a set of non-linear DDE:
\begin{align}\label{e:equilnl}
-\eta \dot l(t)&=k_1\left(\frac{l(t)-l_0}{l_0}\right)+k_2\left(\frac{l(t)-L(t)}{L(t)}\right),\\
\dot L(t)&=\gamma\left(l(t-\tau)-L(t-\tau)\right).
\end{align}

We aim at studying the oscillatory character and stability of these equations. However, due to their non-linearity we cannot directly apply the methodology previously presented. We aim instead at analysing the linearised form of equation \eqref{e:equilnl} at time $t_0$. By setting $\delta l(t)=l(t)-l(t_0)$ and $\delta L(t)=L(t)-L(t_0)$, the linear terms read,
\begin{align}\label{e:nonl}
-\eta\delta\dot l(t)=&\frac{k_1}{l_0}\delta l+\frac{k_2}{L(t_0)}\delta l(t)-\frac{k_2 l(t_0)}{L(t_0)^2}\delta L(t).
\end{align}

It then follows that by defining the modified stiffness parameters,
\begin{align}
\hat k_1&=\frac{k_1}{l(t_0)} + \frac{k_2}{L(t_0)}\left(1-\frac{l(t_0)}{L(t_0)}\right),\\
\hat k_2&=\frac{k_2 l(t_0)}{L(t_0)^2},   
\end{align}
equation \eqref{e:nonl} is equivalent to the linear terms in the equilibrium equation in \eqref{e:meq}, but replacing $(k_1, k_2)$ by $(\hat k_1, \hat k_2)$  and in terms of $\delta l(t)$ and $\delta L(t)$ instead of $l(t)$ and $L(t)$. This allows us to understand some of the numerical solutions obtained for the non-linear case. 

\begin{figure}[!htb]
 \centering 
 \captionsetup{justification=centering}
 \centering 
\subfigure[Time evolution of current length and rest-length]{\label{/nonlinear_lengthVStime_sustained}\includegraphics[width=.51\linewidth]{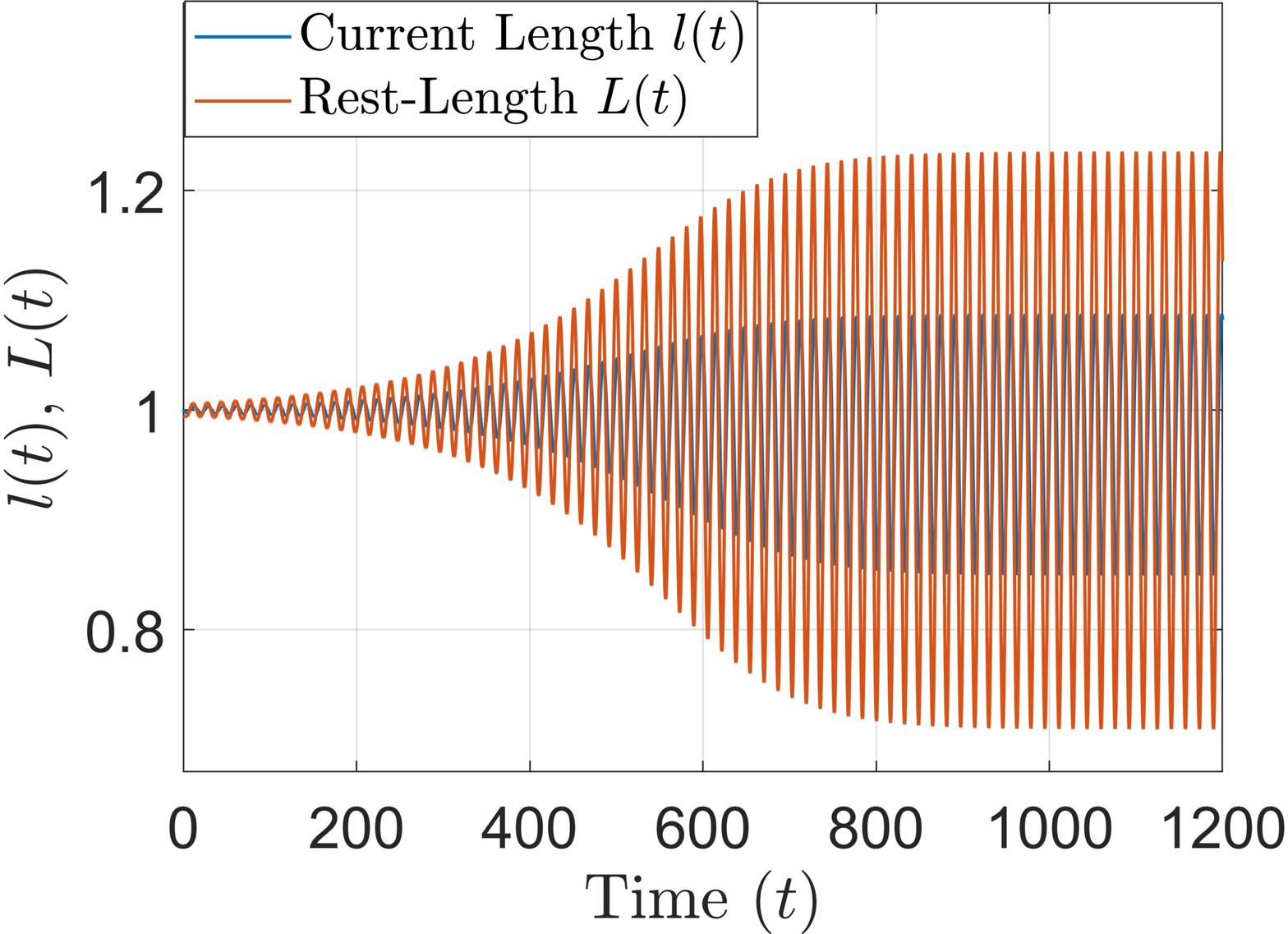}}
\centering
\subfigure[The evolution of $\tilde{k_1}$ and $\tilde{k_2}$.]{\label{nonlinear_k1k2_sustained}\includegraphics[width=0.48\linewidth]{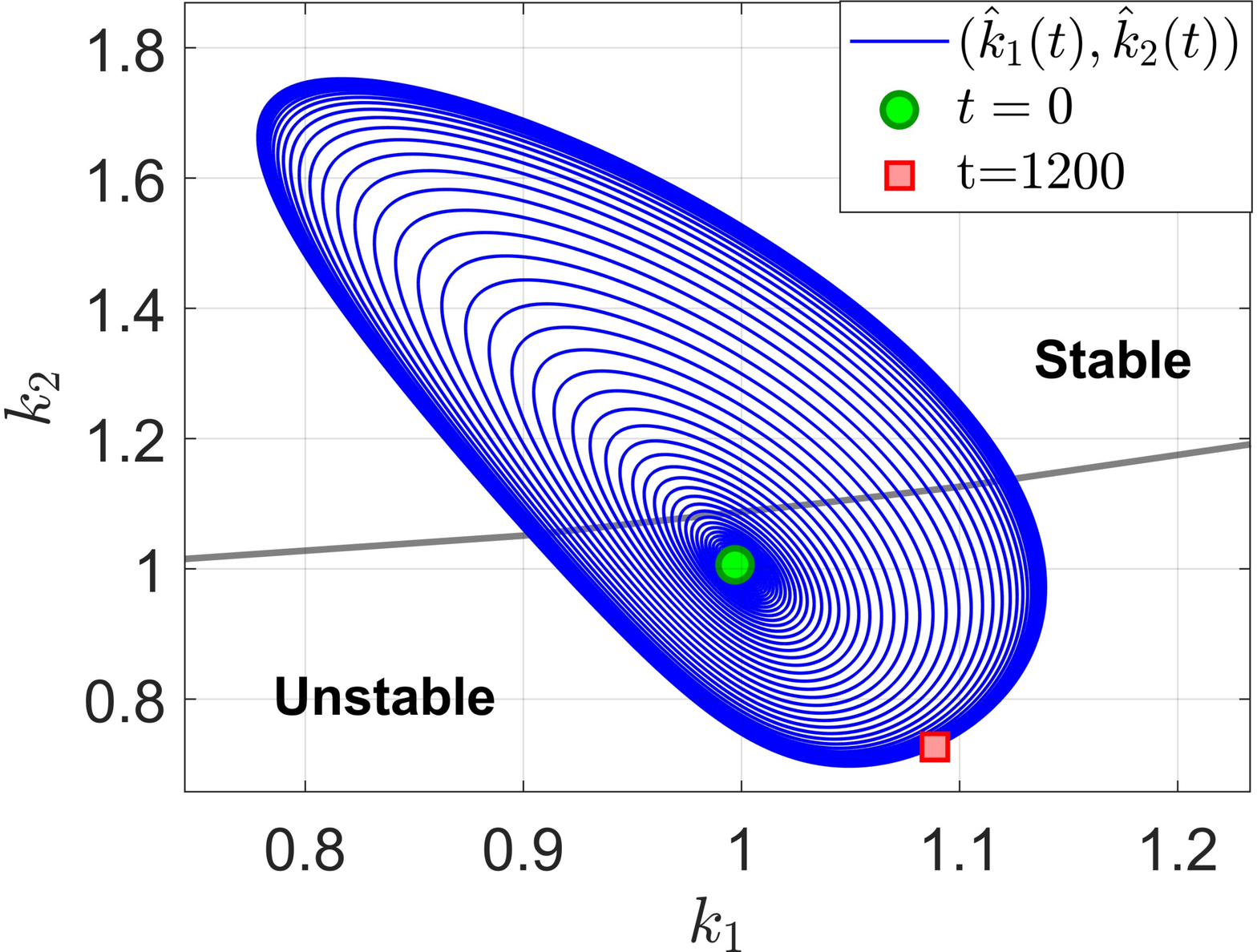}}
\caption{Numerical solution with sustained oscillations of the non-linear model. Parameters: $k_1=1$, $k_2=1$,  $\eta=3$, $\gamma=0.6$, $\tau=5$}
\label{f:noninear_sustained}
\end{figure}

Figure \ref{f:noninear_sustained}a shows the time evolution of $l(t)$ and $L(t)$, which are sustained, that is, their asymptotic behaviour does  not increase nor decrease. We plot in the parametric space of $k_1$ and $k_2$ the modified parameters $\hat k_1$ and $\hat k_2$ for each time $t_0$, as shown in Fig. \ref{f:noninear_sustained}b. It can be observed that although the initial values are located in the unstable region, they in turn oscillate between the unstable and stable region, reaching a limit cycle that alternates between the two domains.

\begin{figure}[H]

 \centering 
 \captionsetup{justification=centering}
 \centering 
\subfigure[Time evolution of current length and rest-length]{\label{NumericalSolutionNonlinearStable}\includegraphics[width=.51\linewidth]{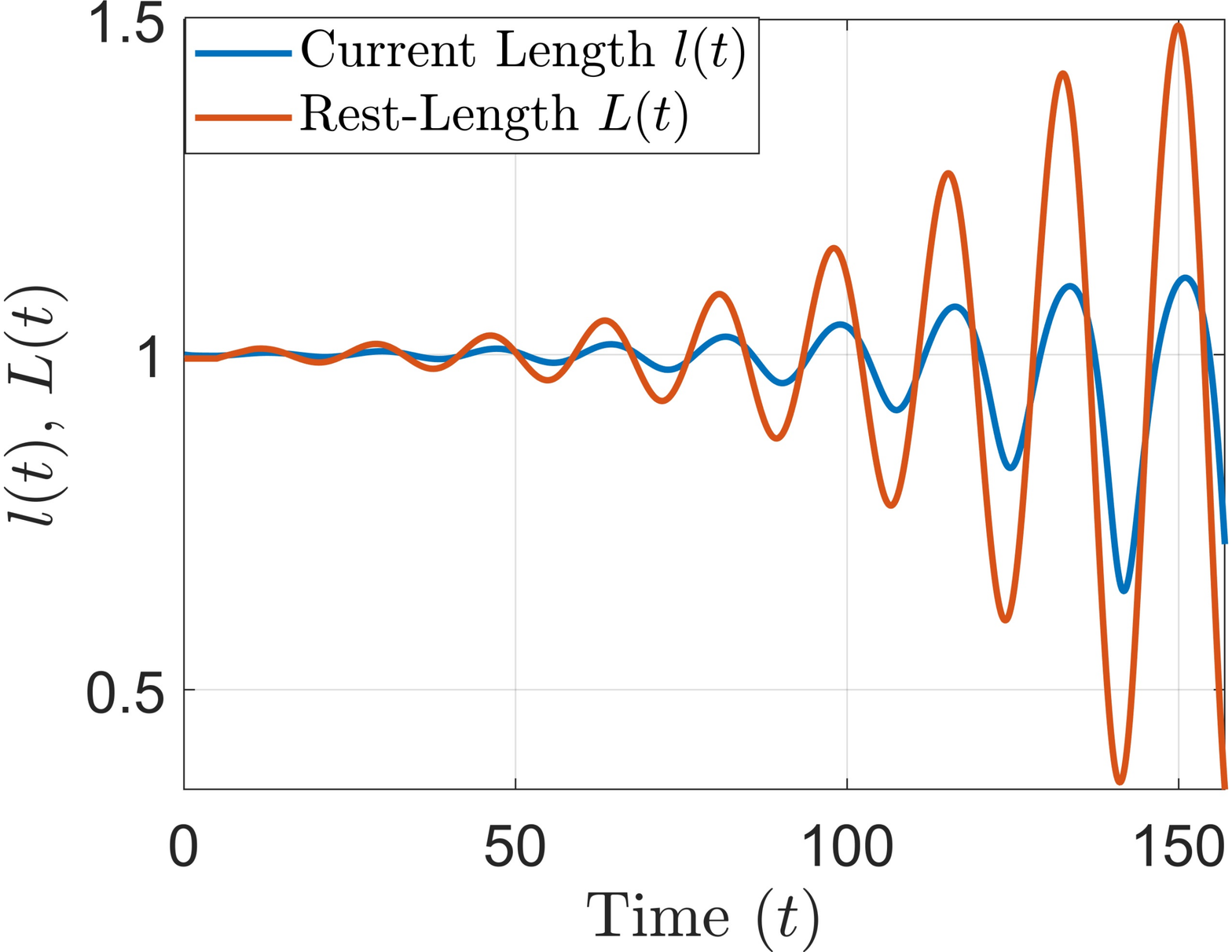}}
\centering
\subfigure[The evolution of $\tilde{k_1}$ and $\tilde{k_2}$]{\label{nonlinear_k1Vsk2_unstable}\includegraphics[width=0.48\linewidth]{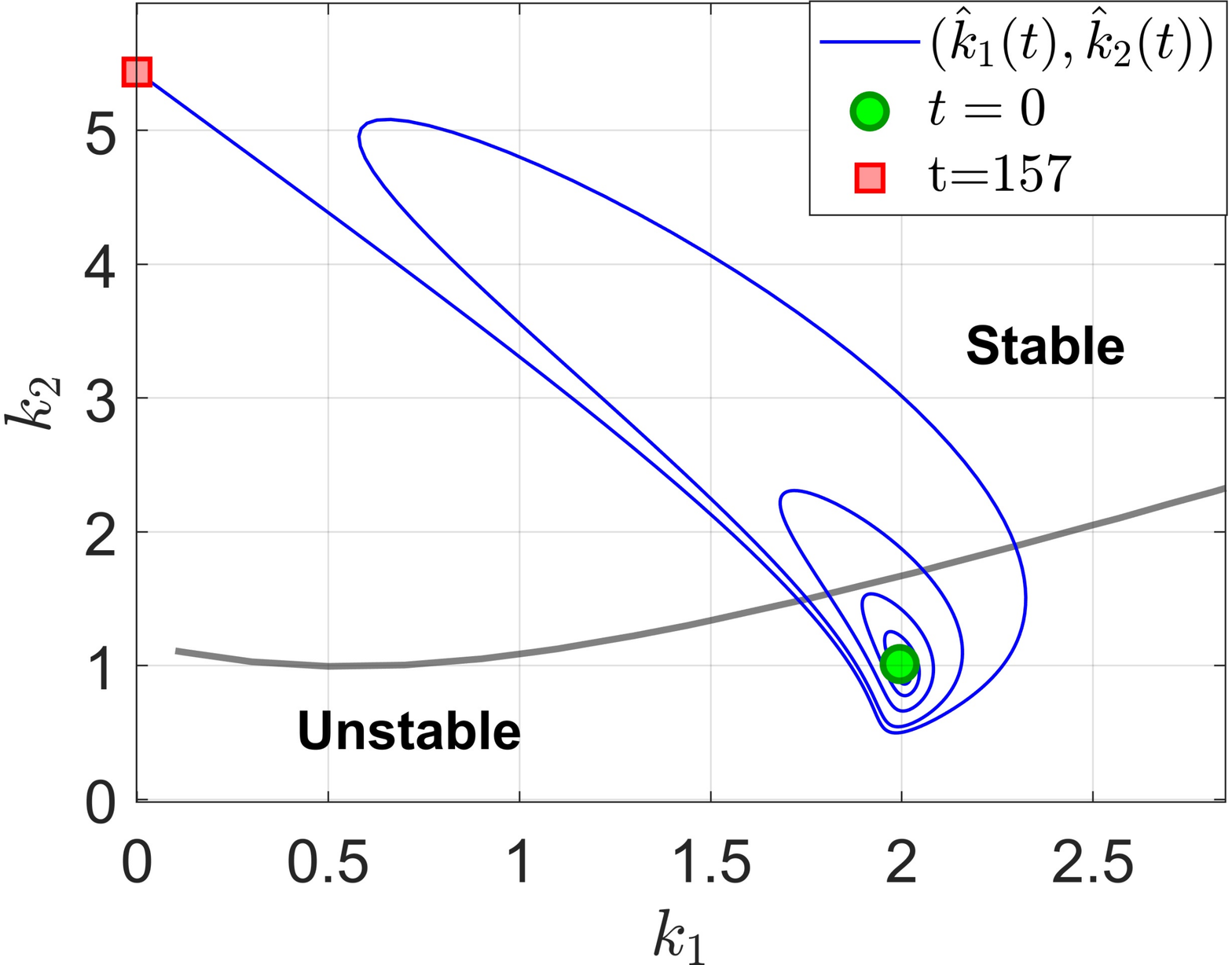}}
\caption{Numerical solution with unstable oscillations on the non-linear model. Parameters: $k_1=2$, $k_2=1$,  $\eta=3$, $\gamma=0.6$, $\tau=5$}
\label{f:nonlinearUnstable}
\end{figure}

\begin{figure}[H]
 \centering 
 \captionsetup{justification=centering}
 \centering 
\subfigure[Time evolution of current length and rest-length]{\label{nonlinear_lengthVStime_stable}\includegraphics[width=.50\linewidth]{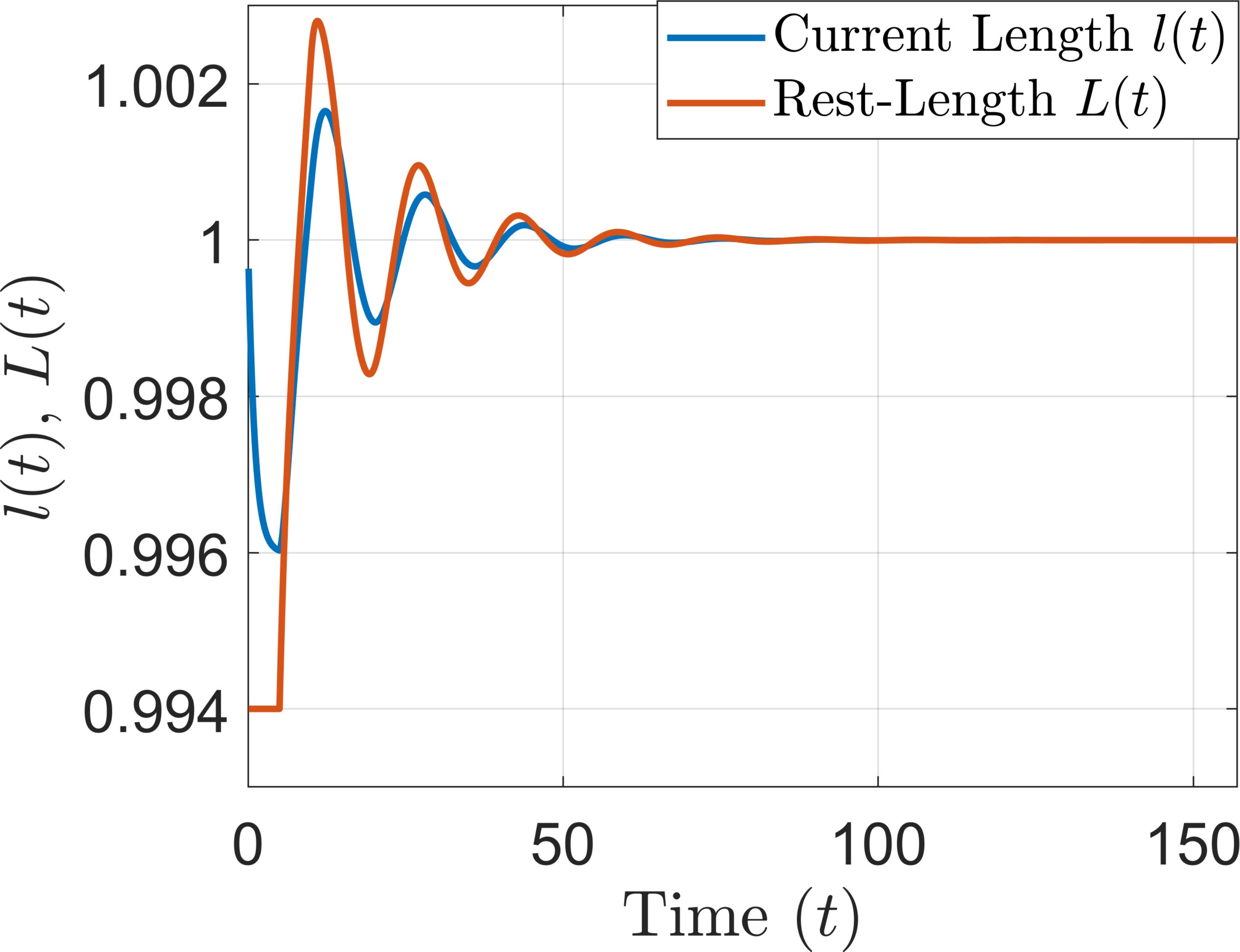}}
\centering
\subfigure[The evolution of $\tilde{k_1}$ and $\tilde{k_2}$]{\label{/nonlinear_k1Vsk2_stable}\includegraphics[width=0.49\linewidth]{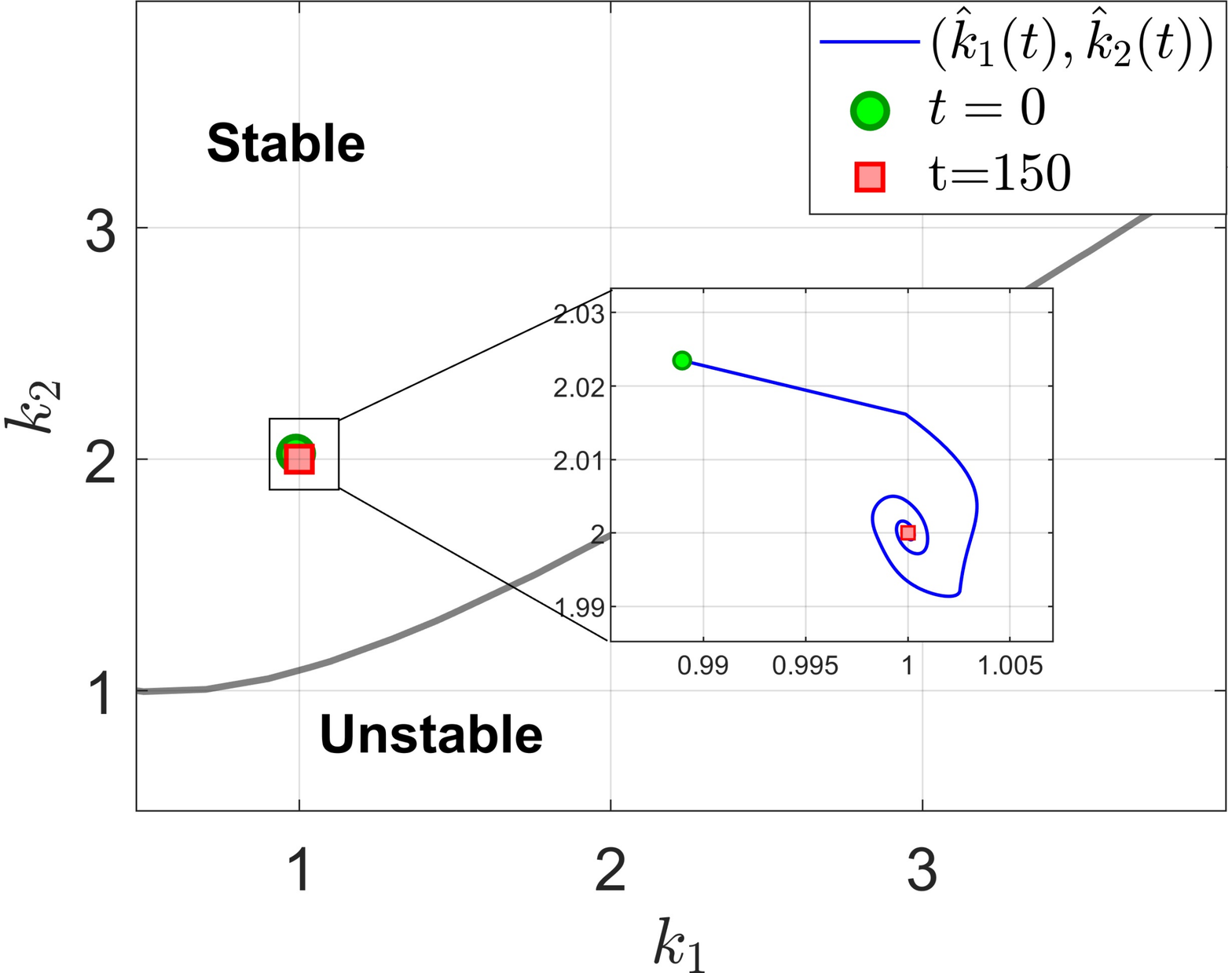}}
\caption{Numerical solution with stable oscillations of the non-linear model. Parameter $k_1=1$, $k_2=2$,  $\eta=3$, $\gamma=0.6$, $\tau=5$}
\label{f:nonlinearStable}
\end{figure}

We have also tested other parameter settings, with an initial location of ($\hat k_1, \hat k_2$) in the parametric space farther from the stability boundary (see Fig. \ref{f:nonlinearUnstable}). In this case, the system exhibits oscillations that reach the singular value $L(t)=0$ for some $t>0$, which renders the DDEs in \eqref{e:equilnl} ill-posed. Instead, when using values that are farther inside the stability region, as it is the case in Fig. \ref{f:nonlinearStable}, the oscillations stabilise before reaching this singular value. Although we are not able to furnish bounds for non-linear stability, we can explain the presence of stable, sustained, or unstable (or singular) oscillations according to the distance of the initial value of $(\hat k_1, \hat k_2)$ to the stability boundary of the linear case.

\section{Conclusions}\label{s:conclusion}

Motivated by the presence of delays and visco-elastic response of tissues, we have presented a rheological model that includes elastic and viscous contributions, and also exhibits oscillatory behaviour.

We have analysed the stability of he model when using a linear strain measure and as a function of the model parameters. We have recovered previous results, which show that increasing values of the delay $\tau$ and the remodelling rate $\gamma$ (a quantity that is inversely proportional to tissue viscosity), render the oscillations unstable.  Remarkably, increasing values of the viscous friction of the domain with respect to external boundary also destabilise the system.

By studying the characteristic function of the DDE we have provided sufficient conditions of stability and bounds to the stability region. This analysis have also allowed us to explain the presence of sustained oscillations in a non-linear version of the model. This persistent oscillations in the tissue deformations  are frequently observed \cite{petrolli19,peyret18}, and in our model are due to the transition between stable and unstable domains. 

We note that despite visco-elastic models based on rest-length changes are increasingly common \cite{cavanaugh20,clement17,khalilgharibi19}, their stability in the presence of delayed response has not been studied. We here 
provide such an analysis which may also help to explain the observed sudden deformations in embryo development and morphogenesis. 
%

\section*{acknowledgements}
JJM and MD have been financially supported by the Spanish Ministry of Science, Innovation and Universities  (MICINN) with grant DPI2016-74929-R  and by the local government \emph{Generalitat de Catalunya} with grant   2017 SGR 1278.


\bibliographystyle{plain}      


\end{document}